\documentclass[10pt,letterpaper,twocolumn]{article} %% two column, final layout

\usepackage{ol2}
\usepackage{amsmath}
\usepackage{amsfonts}
\usepackage{amssymb}
\usepackage[draft]{hyperref}

\begin{document}

\twocolumn[ %% activate for two-column option

\title{Ramsey spectroscopy with squeezed light}

%% For REVTeX it is possible to automate superscript and e-mail callouts with the superscriptaddress option; see REVTeX4 documentation.

\author{Kenan Qu$^*$ and G. S. Agarwal }

\address{Department of Physics, Oklahoma State University, Stillwater, Oklahoma - 74078, USA \\
$^*$Corresponding author: k.qu@okstate.edu
}

\begin{abstract}
Traditional Ramsey spectroscopy has the frequency resolution $2\pi/T$, where $T$ is the time separation between two light fields. Using squeezed states and two-atom excitation joint detection, we present a new scheme achieving a higher resolution $\pi/T$. We use two mode squeezed light which exhibits strong entanglement.
\end{abstract}


 ] %% activate for two-column option

\noindent
Ramsey's method~\cite{qu-Ramsey}  of using two spatially separated optical fields has stood as one of the most successful spectroscopic techniques. It was extended from the radio frequency domain to the optical domain~\cite{qu-Baklanov} in 1970s, and was experimentally demonstrated~\cite{qu-Bergquist,qu-Cohen-Tannoudji} shortly afterwards. As the advantages of the entanglement and squeezing played more and more important role in improving the precision of measurements, Wineland et al proposed~\cite{qu-Wineland-PRA} the application of atomic squeezed spin states to the reduction of quantum noise in Ramsey spectroscopy. Further, the resolution at Heisenberg limit was achieved~\cite{qu-Wineland-Science} using atoms prepared in Greenberger-Horne-Zeilinger states. On the other hand, Agarwal and Scully~\cite{qu-GSA-MOS1} showed that the classical light field in the Ramsey setup can be replaced by nonclassical one leading to the enhancement of the signal-to-noise ratio. Moreover, broadband ultrashort laser pulses~\cite{Eikema} have been used in a Ramsey like setup to improve the resolution. The Ramsey technique has been combined with coherent population trapping techniques to probe the clock transitions with much higher accuracy~\cite{Zanon}, one as can surpass the limits set by collisions and saturation effects. In view of the widespread applications of the Ramsey spectroscopy, we propose a Ramsey spectroscopic scheme using two-atom excitation joint detection and squeezed light field. We are able to achieve a higher resolution using squeezed light rather than using entangled atoms or spin squeezing. We note that the advantages of using squeezed light in optical interferometers are well-known~\cite{Dowling,Gerry,Caves}.

\begin{figure}[hpt]
\centerline{\includegraphics[width=0.43\textwidth]{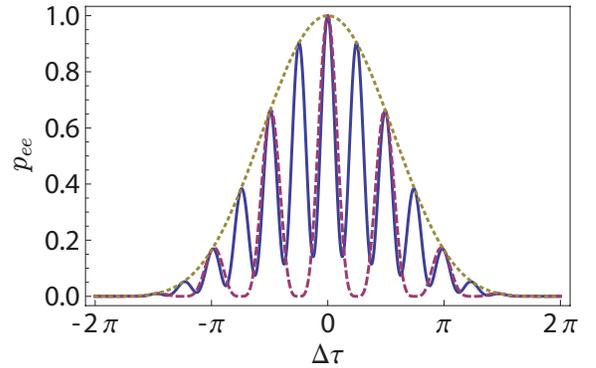}}
 \caption{\label{Fig-3} The interference fringes in two-atom excitation joint detection: solid curve, squeezed light excitation with $r=0.3$ and $T=4\tau$; dashed curve, coherent state excitation. The dotted curve denotes $p_{ee}$ for single zone coherent excitation probability. All the plots are normalized to their maximum values. }
\end{figure}

We start by recalling the method to measure the resonance frequency of a two level atom. In a traditional spectroscopy, an atom, prepared in ground state, passes through a light field region during time $\tau$. The probability of detecting it in the excited state has the form $\tilde{p}_e=g^2\tau^2\mathrm{sinc}^2(\Delta\tau/2)$ where $\Delta$ is the frequency detuning between the atomic resonance frequency and the light field. Its resolution is restricted by the transit-time broadening. The most successful technique solving this problem is Ramsey's ingenious idea of using the interference arising from the interaction of a quantum system with two regions of phase-coherent fields separated by time interval $T$. The time $T$ can be much longer than $\tau$ as long as it remains smaller than the atomic dephasing time. The probability of detecting one atom getting excited after passing two light field zones exhibits the Ramsey fringe $p_e=\tilde{p}_e\cdot(\cos\Delta T+1)$. Considering that the spatial separation of these two light field zones can be much larger than the width of each one, Ramsey fringe has much sharper peaks and thus improves the resolution for the atomic frequency by a factor of $T/\tau$, as can be seen in Fig.~\ref{Fig-3}. Ramsey's method is based on the quantum interference between the transition amplitudes in the two excitation zones~\cite{GSAbook}. The Ramsey fringes appear when detecting its excitation probability with respect to different $\tau$ and $T$. The frequency detuning $\Delta$, and hence the atomic resonance frequency, can be obtained as the inverse of the fringe period.

Based on Ramsey's method, we can further improve the fringe resolution by using two-atom excitation joint detection. The probability of two-atom excitation by a coherent field has the form $p_{ee}=\tilde{p}_e^2\cdot(\cos\Delta T+1)^2 = \tilde{p}_e^2\cdot(\frac12\cos2\Delta T+2\cos\Delta T+\frac32)$. This form has a term $\cos2\Delta T$ which oscillates in time domain twice as fast as the single excitation detection probability. However, we can also see that its fringe pattern is a mixture of $\cos2\Delta T$ and $\cos\Delta T$. The slow oscillating term $\cos\Delta T$ comes from the probability that only one atom is excited. If both the two atoms can always be excited simultaneously, the latter term containing the more slowly oscillating can be eliminated, thus one would be left with a pattern resolution improved by a factor of $2$. We show that squeezed light can be used to eliminate the term $\cos\Delta T$ which then leads to the resolution improvement. We hope to discuss the sensitivity of detection elsewhere.

\begin{figure}[htp]
 \centerline{\includegraphics[width=0.4\textwidth]{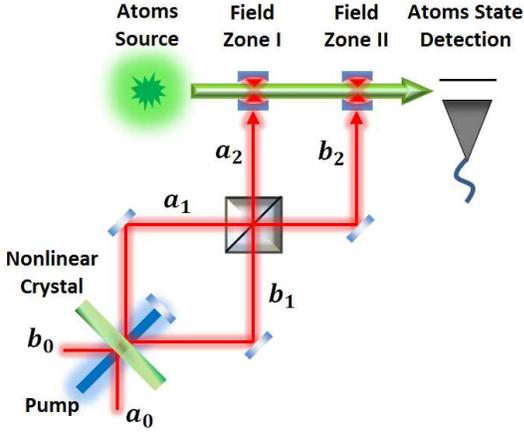}}
 \caption{\label{Fig-2} Schematic of the experimental setup for the Ramsey spectroscopy using squeezed light excitation..}
\end{figure}

Consider the setup, as illustrated in Fig.~\ref{Fig-2}, for two-atom excitation by a field in a squeezed state.
The light fields are created by the spontaneous parametric down-converter (SPDC), and then incident on a symmetric lossless beam splitter (BS). These feed the two zones in which atoms are excited. We identify the two input ports as $a_j$ and $b_j$, $j=1,2$. Here $a_j$ and $b_j$ are the annihilation operators for the two modes. The light fields prepared in the field zones at time $t=0$ are $|\varphi(0)\rangle = \hat{U}\hat{S}(\xi)|0,0\rangle$, where $\hat{U}=\exp[i\theta(a^\dag b+b^\dag a)]$ is the beam splitter operator and $\hat{S}(\xi)=\exp(\xi a^\dag b^\dag - \xi^*ab)$ is the two-mode squeezing operator with $\xi=r\mathrm{e}^{i\varphi}$. Although the two-mode squeezed states are entangled states, they are separable when they pass through the BS and they can be written as
\begin{equation}\label{1}
    |\varphi(0)\rangle = \exp[\frac12(\xi a_2^{\dag2} - \xi^*a_2^2)]|0\rangle \exp[\frac12(\xi b_2^{\dag2} - \xi^* b_2^2)]|0\rangle.
\end{equation}
We see that each light field is a squeezed vacuum state. The vacuum input states obey the commutation relation $[a_0,a_0^\dag]=[b_0,b_0^\dag]=1$ and $[a_0,b_0]=0$. It is straightforward to prove that $a_2$ with $b_2$, $a_1$ with $b_1$ and $a_0$ with $b_0$ bear the same commutation relation. The atomic excitation in the zones I and II takes place via uncorrelated fields, although the properties of the fields $a_2$ and $b_2$ depend on the entanglement originally present between the fields $a_1$ and $b_1$. We will see later in (\ref{31}) that this is the reason the $\cos\Delta T$ term does not appear in the excitation probability.

In the two-atom joint count configuration, we prepare two atoms in ground state $|g_i,g_j\rangle$, and the atoms-field system is in the state $|\psi(0)\rangle = |g_i,g_j\rangle|\varphi(0)\rangle$.  The atoms are sent through the two field zones, and afterwards, collected by the detector. The detector registers if both the two atoms are detected in the excited state. Ramsey fringes appear in the signal registered by the detector when one scans $\Delta T$ which is controlled by the optical field frequency and the pulses separation in time domain. Next we calculate the atom excitation probabilities. When the atoms pass through the light field zones, they interact with the light fields and the excitation process is described by the Hamiltonian in the interaction picture~\cite{qu-GSA-MOS1},
\begin{multline}\label{2}
    H_1(t) = \sum_i \hbar(g_iS_i^\dag a_2\mathrm{e}^{\mathrm{i}\Delta t} + h.c.)\theta(\tau-t) \\
        + \sum_i \hbar(g_iS_i^\dag b_2\mathrm{e}^{\mathrm{i}\Delta t} + h.c.)\theta(t-T-\tau)\theta(T+2\tau-t).
\end{multline}
The step function $\theta(t)$ represents the time regions of interaction in the two field zones. The $g_i$ is the coupling constants. Notice the same phase factor $\mathrm{e}^{\mathrm{i}\Delta t}$ indicates that the two fields have the same phase. The final state for the atoms-field system can be calculated by using the solution of the Schr\"{o}dinger equation up to second order in the coupling constants,
\begin{align}\label{ec}
    & |\psi(t)\rangle \simeq |g_i,g_j\rangle|\varphi(0)\rangle & \nonumber \\
    & \quad  -i f(\tau) g(a_2+b_2\mathrm{e}^{i\Delta T})[|e_i,g_j\rangle|\varphi(0)\rangle + |g_i,e_j\rangle|\varphi(0)\rangle] \nonumber \\
    & \quad + \left[ -i f(\tau) g(a_2+b_2\mathrm{e}^{i\Delta T}) \right]^2 |e_i,e_j\rangle|\varphi(0)\rangle.
\end{align}
where $f(\tau) = \frac{\mathrm{e}^{i\Delta\tau}-1}{i\Delta}$. In the limit $T\gg\tau$, the single-atom and two-atom excitation  probabilities after passing the two fields can be calculated
\begin{align}
    p_{e} &= \mathrm{Tr_{field}} \langle e_i,g_j|\psi(t)\rangle\langle\psi(t)|e_i,g_j\rangle \nonumber \\
    &= \tilde{p}_e\langle (a_2^\dag + b_2^\dag\mathrm{e}^{-\mathrm{i}\Delta T})(a_2 + b_2\mathrm{e}^{\mathrm{i}\Delta T}) \rangle  \nonumber \\
    &= 2\tilde{p}_e\sinh^2r, \label{31} \\
    p_{ee} &= \mathrm{Tr_{field}} \langle e_i,e_j|\psi(t)\rangle\langle\psi(t)|e_i,e_j\rangle \nonumber \\
    &= \tilde{p}_e^2\langle (a_2^\dag + b_2^\dag\mathrm{e}^{-\mathrm{i}\Delta T})^2(a_2 + b_2\mathrm{e}^{\mathrm{i}\Delta T})^2 \rangle \nonumber \\
    &= \tilde{p}_e^2[\frac12 \sinh^2 2r\cos(2\Delta T) + 8\sinh^4 r + \frac12 \sinh^2 2r] \nonumber \\
    &= \tilde{p}_e^2 [1 + V\cos(2\Delta T)](8\sinh^4 r + \frac12 \sinh^2 2r). \label{4}
\end{align}
where $\tilde{p}_e = f^2(\tau)g^2 = g^2\tau^2\mathrm{sinc}^2(\Delta\tau/2)$ and $V = 1/(1 + 4\tanh^2r)$. The visibility of the fringes will be given by $V$. In the derivation, we used $\langle a_2^{\dag2}a_2^2\rangle = \langle b_2^{\dag2}b_2^2\rangle = \sinh^2r+3\sinh^4r$, $\langle a_2^{\dag}a_2\rangle\langle b_2^{\dag}b_2\rangle=\sinh^4r$, $\langle a_2^{\dag2}b_2^2\rangle = \langle b_2^{\dag2}a_2^{\dag2}\rangle = \sinh^2r\cosh^2r$, and $\langle a_2^{\dag}b_2\rangle = \dots =0$. These properties depend on the presence of entanglement between the fields $a_1$ and $b_1$.

\begin{figure}[htp]
\centerline{\includegraphics[width=0.4\textwidth]{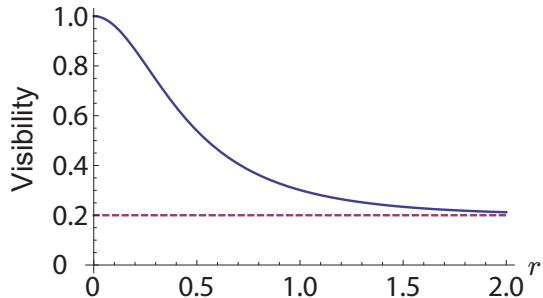}}
 \caption{\label{Fig-4} The contrast of the interference fringes in the two-atom excitation counts with the squeezing parameter $r$.}
\end{figure}

The two-atom excitation count with coherent state produces Ramsey fringes $p_{ee}=\tilde{p}_e^2\cdot(\frac12\cos2\Delta T+2\cos\Delta T+\frac32)$. These are dominated by the term $2\cos\Delta T$ which has a resolution $2\pi/\Delta T$. The resolution is defined as the separation between two adjacent maxima. The expression (\ref{4}) shows that the Ramsey fringes under squeezed state excitation reveal an interference of the form $\cos(2\Delta T)$, without the slowly oscillating term $\cos\Delta T$. We illustrate $p_{ee}$ for both coherent state input (dashed curve) and the squeezed vacuum input (solid curve) in Fig.~\ref{Fig-3}, and we can see that, in contrast to the coherent state input, the squeezed vacuum state input produces Ramsey fringes with only fast oscillating terms with resolution $\pi/\Delta T$. This enables one to determine the atomic transition frequency with higher precision. The visibility of the fringes depends on the squeezing parameter $r$ which also quantifies the entanglement in the two-mode squeezed vacuum (Eq.(3.52) in~\cite{GSAbook}). As shown in Fig.~\ref{Fig-4}, it drops with increase in the squeezing parameter. A similar result is known in quantum lithography~\cite{OPA,DowlingExp}.

To understand the physics behind the improvement, we recall that the Ramsey fringes arise from the lack of knowledge in which path the atoms get excited. In the Ramsey spectroscopy with coherent light, the light fields are not changed after the atoms are excited, since subtraction of a photon from a coherent state leaves the state unchanged. Therefore we have no way to know in which path the atoms get excited and, as a result, we obtain the Ramsey fringes. However, with squeezed light containing a photon distribution with only even photon numbers, the parity of the light field is changed from even to odd once a single atom is excited(~\cite{Namekata}, and Sec. 4.6 in~\cite{GSAbook}). In another word, we would be able to tell the path in which the single atom gets excited by measuring the parity of the light fields after the atom passes through, which eliminates the interference term $\cos\Delta T$ term in the Ramsey fringes corresponding to a single-atom excitation probability. Keep in mind that the loss of interference term $\cos\Delta T$ is not due to the random phase fluctuation of the light fields, but due to the availability of the which-path information. We note that the structure of $p_{ee}$ has a nice interpretation which follows from the last term in (3). The term $a_2^2(b_2^2)$ would correspond to the excitation in zone I(II). These two two-atom excitation  amplitudes interfere leading to $\cos(2\Delta T)$ term in (5). The two-atom excitation amplitude resulting from single excitation in each zone leads to the background term in (5). The interferences arise because the single mode squeezed vacuum has the important property that $\langle a^2\rangle\neq0$, even though $\langle a\rangle=0$. This reminiscent of the interference between two independent sources~\cite{Walther,Mandel} which can occur at the level of intensity-intensity correlations. In our case, this is the quantity $\langle :II:\rangle$, with $I=(a_2^\dag+b_2^\dag\mathrm{e}^{-\mathrm{i}\Delta T})(a_2 + b_2\mathrm{e}^{\mathrm{i}\Delta T})$.

Finally, it should be noted that in the experiments, one can use a different configuration by applying two light pulses with time interval $T$ onto an ensemble containing a low density (so that collisions are unimportant) of atoms, and monitoring the fluorescence light from the excited atoms in the ensemble~\cite{Eikema,Zanon}. The fluctuations in fluorescence are determined by the quantity $(p_{ee}-p_e^2)$ and hence such fluctuations would provide a more accurate method for determining the atomic frequencies.

In conclusion, we have shown how Ramsey spectroscopy with squeezed light beams could provide us a method with increased resolution. Our results suggest the usefulness of nonclassical light beams in Ramsey spectroscopy. Our work can be extended to the case of the two photon Ramsey spectroscopy, the results for which would depend on the higher order quantum correlations of the field~\cite{Kimble}.

%\pagebreak

\end{document}